\documentstyle[aasms4]{article}
\newcommand{\beq}{\begin{equation}}
\newcommand{\eeq}{\end{equation}}
\newcommand{\etal}{{\sl et~al.~}}

\def\fdg{\hbox{$.\!\!^\circ$}}

\begin{document}

\title{Astrometry with Hubble Space Telescope:
A Parallax of the Central Star of the Planetary Nebula NGC 6853\footnote{Based on 
observations made with
the NASA/ESA Hubble Space Telescope, obtained at the Space Telescope
Science Institute, which is operated by the
Association of Universities for Research in Astronomy, Inc., under NASA
contract NAS5-26555}}

\author{ G.\ Fritz Benedict\altaffilmark{1}, B. E.
McArthur\altaffilmark{1}, L.W.\
Fredrick\altaffilmark{12}, T. E. Harrison\altaffilmark{13}, M. F. 
Skrutskie\altaffilmark{12}, C. L. Slesnick\altaffilmark{12}, J. 
Rhee\altaffilmark{12}, R. J. Patterson\altaffilmark{12},
E.\ Nelan\altaffilmark{5}, W.\ H.\ Jefferys\altaffilmark{6},  
W.~van~Altena\altaffilmark{7}, T. Montemayor\altaffilmark{1}, 
P.~J.~Shelus\altaffilmark{1}, O. G. Franz\altaffilmark{2}, L.\ 
H. Wasserman\altaffilmark{2},
 P.D. Hemenway\altaffilmark{8}, R. L.
Duncombe\altaffilmark{9}, D. Story\altaffilmark{10}, A.\ L.\
Whipple\altaffilmark{10}, and A. J. Bradley\altaffilmark{11} }

\altaffiltext{1}{McDonald Observatory, University of Texas, Austin, TX 78712}
\altaffiltext{2}{Lowell Observatory, 1400 West Mars Hill Rd., Flagstaff, AZ 86001}
\altaffiltext{5}{Space Telescope Science Institute, 3700 San Martin Dr., 
Baltimore, MD 21218}
\altaffiltext{6}{ Department of Astronomy, University of Texas, Austin, TX 78712}
\altaffiltext{7}{ Department of Astronomy, Yale University, PO Box 208101, New 
Haven, CT 06520}
 \altaffiltext{8}{Department of Oceanography, University of Rhode Island, 
Kingston, RI 02881}
\altaffiltext{9}{Department of Aerospace Engineering, University of Texas, Austin, 
TX 78712}
\altaffiltext{10}{ Jackson and Tull, Aerospace Engineering Division
7375 Executive Place, Suite 200, Seabrook, Md.  20706}
\altaffiltext{11}{Spacecraft System Engineering Services, PO Box 91, Annapolis 
Junction, MD 20706}
\altaffiltext{12}{ Department of Astronomy, University of Virginia, PO Box 3818, 
Charlottesville, VA 22903}
\altaffiltext{13}{Department of Astronomy, New Mexico State University, Las 
Cruces, New Mexico 88003}

% Notice that each of these authors has alternate affiliations, which
% are identified by the \altaffilmark after each name.  The actual alternate
% affiliation information is typeset in footnotes at the bottom of the
% first page, and the text itself is specified in \altaffiltext commands.
% There is a separate \altaffiltext for each alternate affiliation
% indicated above.

% The abstract environment prints out the receipt and acceptance dates
% if they are relevant for the journal style.  For the aasms style, they
% will print out as horizontal rules for the editorial staff to type
% on, so long as the author does not include \received and \accepted
% commands.  This should not be done, since \received and \accepted dates
% are not known to the author.

\begin{abstract}
We present an absolute parallax and relative proper motion for the central star of 
the planetary nebula NGC 6853 (The Dumbell). We obtain these with astrometric data 
from
FGS 3, a white-light interferometer on {\it HST}. Spectral classifications and 
VRIJHKT$_2$M and DDO51 photometry of the stars comprising the astrometric 
reference frame provide spectrophotometric estimates of their absolute parallaxes. 
Introducing these into our model as observations with error, we find $\pi_{abs} = 
2.10 \pm 0.48$ mas for the DAO central star of NGC 6853. A weighted average with a 
previous ground-based USNO determination yields $\pi_{abs} = 2.40 \pm 0.32$. We 
assume that the extinction suffered by the reference stars nearest (in angular 
separation and distance) to the central star is the same as for the central star. 
Correcting for color differences, we find $<$A$_V>$ = 0.30 $ \pm $ 0.06 for the 
central star, hence, an absolute magnitude M$_V = 5.48^{-0.16}_{+0.15}$. A recent 
determination of the central star effective temperature aided in estimating the 
central star radius,
$R_{*}=0.055 \pm 0.02R_{\sun}$, a star that may be descending to the white dwarf 
cooling track.

\end{abstract}

% The different journals have different requirements for keywords.  The
% keywords.apj file, found on aas.org in the pubs/aastex-misc directory,
% contains a list of keywords used with the \apj and Letters.  These are
% usually assigned by the editor, but authors may include them in their
% manuscripts if they wish.

\keywords{astrometry --- interferometry --- stars: distances --- stars: white 
dwarf --- planetary nebulae }

% That's it for the front matter.  On to the main body of the paper.
% We'll only put in tutorial remarks at the beginning of each section
% so you can see entire sections together.

% In the first two sections, you should notice the use of the LaTeX \cite
% command to identify citations.  The citations are tied to the
% reference list via symbolic KEYs.  We have chosen the first three
% characters of the first author's name plus the last two numeral of the
% year of publication.  The corresponding reference has a \bibitem
% command in the reference list below.
%
% Please see the AASTeX manual for a more complete discussion on how 
%to make \cite-\bibitem work for you.

\section{Introduction}

Planetary nebulae are a visually spectacular and relatively short-lived step in the evolution from asymptotic giant branch (AGB) stars, to a final white dwarf stage. Iben \& Renzini (1983) first showed that the ejection of most of the gaseous envelope in asymptotic giant branch (AGB) stars occurs at the tip of the thermal pulse phase, in the form of a massive, low-velocity wind. As summarized by
Stanghellini \etal (2002),  the remnant central star (CS) ionizes the gaseous ejecta, while a fast, low mass-loss rate CS wind shapes the PN. PN morphology depends on a complicated combination of phenomena, some occurring within the nebular gas, which evolves in dynamic timescale, and others caused by the evolution of the stellar progenitors and of the CS. Morphology may also depend on the physical status of the interstellar environment of the PN progenitor. Intercomparison of PN can aid our understanding of the complicated astrophysics of this stage of stellar evolution, particularly if distances are known. Many indirect methods of PN distance determination exist (\cite{Cia99} and \cite{Nap01}). Agreement among these methods is seldom better than 20\%. Direct parallax measurments of PN central stars rarely have precisions smaller than the measured parallax, a notable exception being \cite{Har97}, who provide $\sim$ 0.5 mas precision parallaxes for 7 planetary nebulae CS nearer than 500 pc. 

As the last object on the {\it Hubble Space Telescope } ({\it HST}) Astrometry Science Team list of astrophysically interesting stars,  we have determined the absolute parallax of the CS of NGC 6853 (The Dumbell, M27) using FGS 3. \cite{Nap99} classifies the central star as a white dwarf of type DAO. Our extensive investigation of the astrometric reference stars provides an independent estimation of the line of sight extinction to NGC 6853, a significant contributor to the uncertainty in the absolute magnitude, M$_V$, of its CS. We present the results of extensive spectrophotometry of the astrometric reference stars, required to correct our relative parallax to absolute; briefly discuss data acquisition and analysis; and derive an absolute parallax for the CS of NGC 6853. Finally, from a weighted average of our new result and that of Harris \etal (1997) we calculate an absolute magnitude for the CS and apply it to derive a stellar radius. 

\cite{Bra91} and \cite{Nel01}
provide an overview of the
FGS 3 instrument and \cite{Ben99} and \cite{Ben02a} describe the fringe tracking (POS) mode astrometric capabilities 
of FGS 3, along with the data acquisition and reduction strategies used in the present study. 
We time-tag our data with a modified Julian Date, $mJD = JD - 2444000.5$, and abbreviate millisecond of arc, mas, throughout.

\section{Observations and Data Reduction}  \label{AstRefs}

\begin{figure}
\epsscale{1.0}
\epsscale{0.5}
\plotone{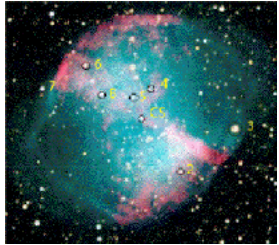}
\caption{NGC 6853 central star and astrometric reference stars. This composite 
color frame maps Johnson B, V, and I bandpasses as blue, green, and red, 
respectively. It illustrates potential nebular effects on reference star 
photometry and spectroscopy. Exposures were obtained with the McDonald Observatory  
Prime Focus Camera on the 0.8m telescope. }
\label{fig-1}
\end{figure}

Figure \ref{fig-1} shows the distribution of the seven reference stars and the CS relative to the brightest regions of the PN. This image, produced by compositing Johnson B, V, and I bandpass frames obtained with the McDonald Observatory 0.8m telescope and Prime Focus Camera, reveals a particular difficulty with these data. The PN emission can contaminate the ancillary photometry and spectroscopy required to generate reference star spectrophotometric parallaxes (Section~\ref{SpecPhot}). Eight sets of astrometric data were acquired with {\it HST}, spanning 2.59 years, for a total of 140 measurements  of  the NGC 6853 CS  and reference stars. Each data set required approximately 33 minutes of spacecraft time. The data were reduced and calibrated as detailed in \cite{Ben02a},  \cite{Ben02b}, and \cite{mca01}. At each epoch we measured reference stars and the target  multiple times, this to correct for intra-orbit drift of the type seen in the cross filter calibration data shown in figure 1 of \cite{Ben02a}. 

Table~\ref{tbl-LOO} lists the eight epochs of observation and highlights another particular difficulty with these data. We obtain observations at each of the two maximum parallax factors; hence the two distinct spacecraft roll values in Table~\ref{tbl-LOO}. These are imposed by the requirement that
{\it HST} roll to keep its solar panels fully illuminated throughout
the year.  This roll constraint generally imposes alternate orientations at each time of maximum positive or negative parallax factor over a typical 2.5 year campaign, usually allowing a clean separation of parallax and proper motion signatures. In this case two guide star acquisition failures on two attempts at roll$\sim 101$\arcdeg~(along with a bookkeeping error in scheduling the makeup observation), left us with the less than satisfactory temporal segregation of orientations shown in Table~\ref{tbl-LOO}. Only the first two epochs occur at maximum negative parallax factor.

\setcounter{footnote}{0}
\section{Spectrophotometric Absolute Parallaxes of the Astrometric Reference Stars} \label{SpecPhot}
Because the parallax determined for the NGC 6853 CS will be
measured with respect to reference frame stars which have their own
parallaxes, we must either apply a statistically derived correction from relative to absolute parallax (Van Altena, Lee \& Hofleit 1995, hereafter YPC95) or estimate the absolute parallaxes of the reference frame stars listed in Table \ref{tbl-POS}. In principle, the colors, spectral type, and luminosity class of a star can be used to estimate the absolute magnitude, M$_V$, and V-band absorption, A$_V$. The absolute parallax is then simply,
\beq
\pi_{abs} = 10^{-(V-M_V+5-A_V)/5}
\eeq

The luminosity class is generally more difficult to estimate than the spectral type (temperature class). However, the derived absolute magnitudes are critically dependent on the luminosity class. As a consequence we obtained additional photometry in an attempt to confirm the luminosity classes. Specifically, we employ the technique used by Majewski et al. (2000) to discriminate between giants and dwarfs for stars later than $\sim$ G5, an approach also discussed by  \cite{Pal94}.

\subsection{Photometry}
Our band passes for reference star photometry include:  BVRI (CCD photometry from a 0.4m telescope at New Mexico State University), JHK (from a pre-release of 2MASS\footnote{The Two Micron All Sky Survey
is a joint project of the University of Massachusetts and the Infrared Processing
and Analysis Center/California Institute of Technology }), and Washington/DDO filters M, DDO51, and T$_2$ (obtained at McDonald Observatory with the 0.8m and Prime Focus Camera).  The 2MASS JHK have been transformed to the Bessell \& Brett (1988) system using the transformations provided in \cite{Car01}. The RI are transformed (Bessell, 1979) to the Johnson system from Kron-Cousins measures. Tables \ref{tbl-VIS} and \ref{tbl-IR} list the visible, infrared, and Washington/DDO photometry for the NGC 6853  reference stars, ref-2 through ref-8.

\subsection{Spectroscopy and Luminosity Class-sensitive Photometry}
The spectra from which we estimated spectral type and luminosity class come from the New Mexico State University Apache 
Peak Observatory\footnote{ The
Apache Point Observatory 3.5 m telescope is owned and operated by
the Astrophysical Research Consortium.}. The dispersion 
was 1.61 \AA/pixel with wavelength coverage 4101 -- 4905 \AA. Classifications used a combination of template matching and line ratios. The brightest targets had about 1500 counts
above sky per pixel, or S/N $\sim$ 40, while the faintest targets had about 400 counts
per pixel (S/N $\sim$ 20). The spectral types for the higher S/N stars are within $\pm$1 subclass. Classifications for the lower S/N stars are $\pm$2
subclasses. Table \ref{tbl-SPP} lists the spectral types and luminosity classes for our reference stars. The estimated classification uncertainties are used to generate the $\sigma_{M_V}$ values in that table. 

\begin{figure}
\epsscale{1.0}
\epsscale{0.75}
\plotone{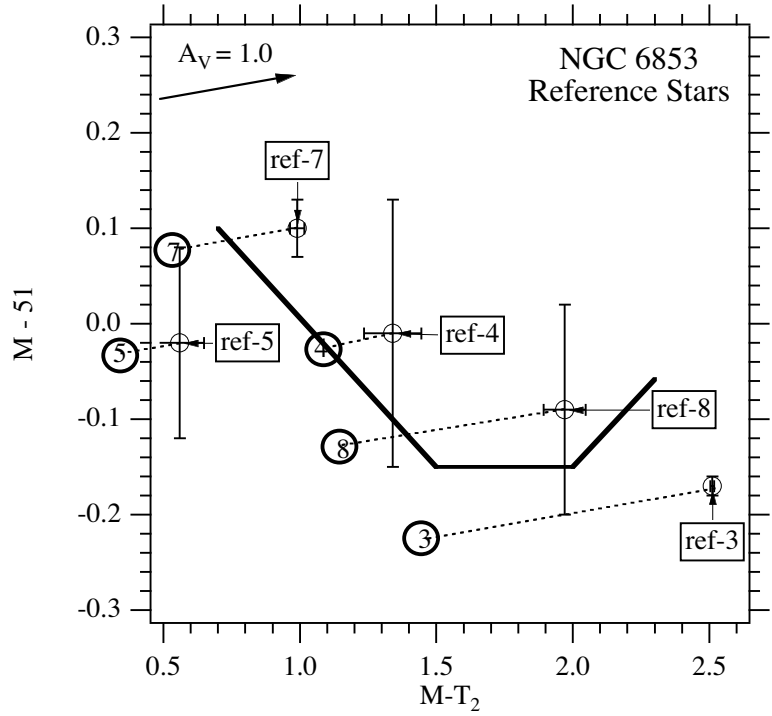}
\caption{M-DDO51 (M-51) vs M-T$_2$ color-color diagram. The solid line is the 
division between luminosity class V and luminosity class III stars. Giants are 
generally above the line, dwarfs below. The reddening vector is for an A$_V$=1.0. 
The circled numbers are the reference star ID's plotted at the de-reddened values, 
based on the per-star $<$A$_V>$ from Table~\ref{tbl-AV}. Nebular emission lines 
will move objects towards bluer M-51 values.}
\label{fig-2}
\end{figure}

\begin{figure}
\epsscale{1.0}
\epsscale{0.65}
\plotone{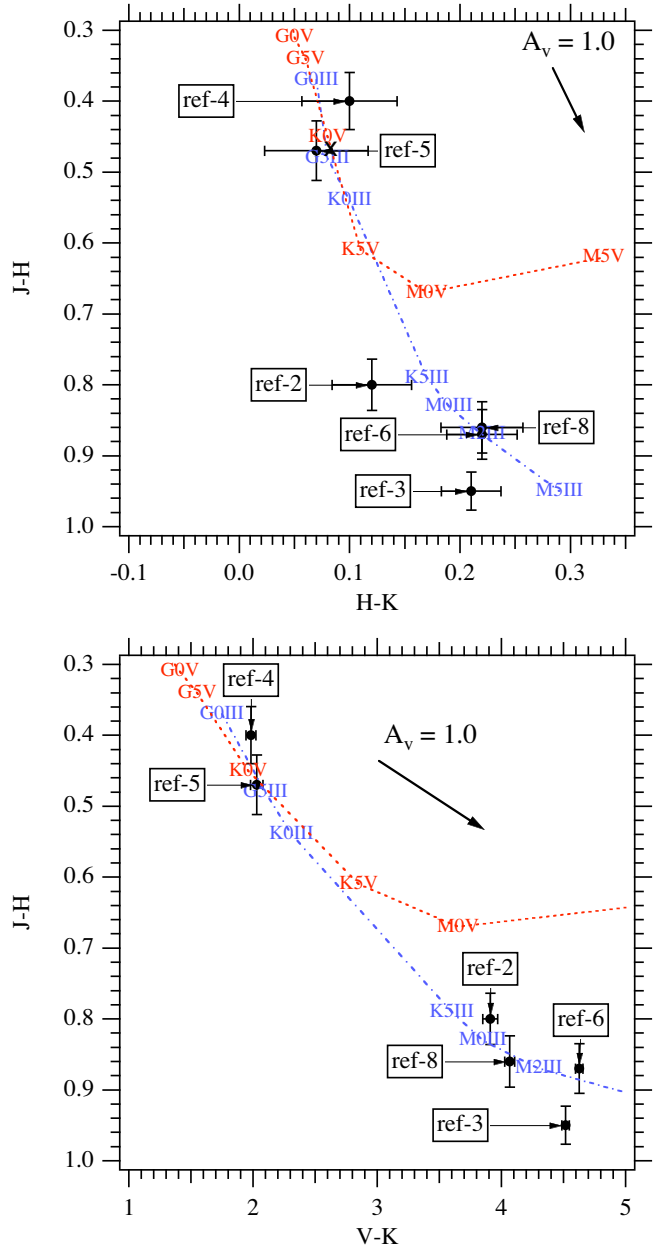}
\caption{J-H vs H-K and J-H vs V-K color-color diagrams. The dashed line is the 
locus of  dwarf (luminosity class V) stars of various spectral types; the dot-
dashed line is for giants (luminosity class III). The reddening vector indicates 
A$_V$=1.0 for the plotted color systems.}
\label{fig-3}
\end{figure}

The Washington/DDO photometry  can provide a possible confirmation of the estimated luminosity class, depending on the spectral type and luminosity class of the star (later than G5 for dwarfs, later than G0 for giants). Washington/DDO photometry is less helpful as a discriminator in this case than it has proved for our previous targets (e.g., Benedict \etal 2002a\nocite{Ben02a}, Benedict \etal 2002b \nocite{Ben02b}, McArthur \etal 2001\nocite{mca01}). As seen in Figure~\ref{fig-1} the nebular emission can 
contaminate the aperture photometry, depending on the filter bandpass. This contamination occurs predominantly in the M filter, because its effective bandpass (4500 - 5400 \AA) includes the strong emission lines at $\lambda$5007 \AA ~[O III] and $\lambda 4861$ \AA ~H$\beta$. This results in broadband M-T$_2$
colors which are slightly bluer than they would be in the absence of the nebula, and significantly bluer
M-DDO51 indices (M-51 in Figure~\ref{fig-2}). Additionally, the nebular emission  
contaminates the sky annulus, which leads to larger measured magnitude errors 
than would be the case in the absence of the nebula.
We list in Table~\ref{tbl-IR} the Washington-DDO photometry. Unfortunately both ref-2 and ref-6 fell on bad columns in the photometer CCD. Figure \ref{fig-3} shows the Washington-DDO photometry along with a dividing line between dwarfs and giants (Paltoglou \& Bell 1994 \nocite{Pal94}). The boundary between giants and dwarfs is actually far 'fuzzier' than suggested by the solid line in Figure \ref{fig-3} and complicated by the photometric transition from dwarfs to giants through subgiants.  This soft boundary is readily apparent in Majewski et al. (2000) figure 14. In the absence of contaminating nebular flux objects just above the heavy line are statistically more likely to be giants than objects just below the line. After correcting for interstellar extinction, our reference stars lie on the dividing line to the left, where giant/dwarf discrimination is poorest. Nebular emission lines could have depressed ref-3 and ref-8 (giants, as determined from spectroscopy) below the giant-dwarf dividing line.

\subsection{Interstellar Extinction} \label{AV}
To determine interstellar extinction we first plot these stars on several color-color diagrams. A comparison of the relationships between spectral type and intrinsic color against those we measured provides an estimate of reddening. Figure \ref{fig-3} contains  J-H vs H-K and J-H vs V-K color-color diagrams and reddening vectors for A$_V$ = 1.0. Also plotted are mappings between spectral type and luminosity class V and III from \cite{Bes88} and \cite{Cox00} (hereafter AQ2000). Figure~\ref{fig-3}, and similar plots for the other measured colors, along with the estimated spectral types, provides an indication of the reddening for each reference star. 

Assuming an R = 3.1 galactic reddening law (Savage \& Mathis 1979\nocite{Sav79}), we derive A$_V$ values by comparing the measured colors (Tables~\ref{tbl-VIS} and \ref{tbl-IR} ) with intrinsic V-R, V-I, J-K, and V-K colors from \cite{Bes88} and AQ2000. Specifically we estimate A$_V$ from four different ratios, each derived from the Savage \& Mathis (1977) reddening law: A$_V$/E(V-R) = 5.1; A$_V$/E(J-K) = 5.8; A$_V$/E(V-K) = 1.1; and A$_V$/E(V-I) = 2.4. We excluded A$_V$/E(B-V) due to the higher errors in that color index. The resulting A$_V$ are collected in Table \ref{tbl-AV}. The errors are the standard deviation of the means for each star. We also tabulate A$_V$ per unit 100 pc distance for each star. Colors and spectral types of the NGC 6853 reference stars are consistent with a field-wide average $\langle$A$_V\rangle$=1.46$\pm$0.35, far less than the maximum reddening, A$_V < 4.66$ determined by \cite{Sch98}.

\subsection{Adopted Reference Frame Absolute Parallaxes}

We derive absolute parallaxes with M$_V$ values from AQ2000 and the $\langle$A$_V\rangle$ derived from the photometry. Our parallax values are listed in Table \ref{tbl-SPP}. Individually, no reference star parallax is better determined than ${\sigma_{\pi}\over \pi}$ = 18\%. The average absolute parallax for the reference frame is $\langle\pi_{abs}\rangle = 1.0$ mas.
As a check we compare this to the correction to absolute parallax discussed and presented
in YPC95 (section 3.2, fig. 2). Entering
YPC95, fig. 2, with the NGC 6853 galactic
latitude, l = -3\fdg7, and average magnitude for the
reference frame, $\langle$ Vref $\rangle$= 14.3, we obtain a correction
to absolute of 1.1 mas. We will use the 1.0 mas correction derived from spectrophotometry. When such data are
available the
use of spectrophotometric parallaxes offers a more direct way of
determining the reference star absolute parallaxes.

\section{Absolute Parallax of the Central Star of NGC 6853}
\subsection{The Astrometric Model}

With the positions measured by FGS 3 we determine the scale, rotation, and offset ``plate
constants" relative to an arbitrarily adopted constraint epoch (the so-called ``master plate") for
each observation set (the data acquired at each epoch). The mJD of each observation set is listed in Table~\ref{tbl-LOO}, along with a measured magnitude transformed from the FGS instrumental system as per \cite{Ben98}. The NGC 6853 reference frame contains 7 stars. We employ the six parameter model discussed in McArthur et al. (2001) for those observations. In this case we determined the plate parameters from reference star data only, then apply them as constants to obtain the parallax and proper motion of the CS. For the NGC 6853 field all the reference stars are redder than the science target. Hence, we apply the corrections for lateral color discussed in Benedict et al. (1999). 

As for all our previous astrometric analyses, we employ GaussFit (\cite{Jef87}) to minimize $\chi^2$. The solved equations
of condition for NGC 6853 are:
\beq
        x' = x + lcx(\it B-V) 
\eeq
\beq
        y' = y + lcy(\it B-V) 
\eeq
\beq
\xi = Ax' + By' + C + R_x (x'^2 + y'^2) - \mu_x \Delta t  - P_\alpha\pi_x
\eeq
\beq
\eta = -Bx' + Ay' + F + R_y(x'^2 + y'^2) - \mu_y \Delta t  - P_\delta\pi_y
\eeq

where $\it x$ and $\it y$ are the measured coordinates from {\it HST};
$\it lcx$ and $\it lcy$ are the
lateral color corrections from Benedict et al. 1999\nocite{Ben99}; and $\it B-V $ are
the  B-V  colors of each star. A  and  B   
are scale and rotation plate constants, C and F are
offsets; $R_x$ and $R_y$ are radial terms;
$\mu_x$ and $\mu_y$ are proper motions; $\Delta$t is the epoch difference from the mean epoch;
$P_\alpha$ and $P_\delta$ are parallax factors;  and $\it \pi_x$ and $\it \pi_y$
 are  the parallaxes in x and y. We obtain the parallax factors from a JPL Earth orbit predictor (\cite{Sta90}), upgraded to version DE405. We imposed the constraint that the reference star proper motions are random in direction by forcing their sum in x and y to be zero, $\sum \mu_x = \sum \mu_y = 0$. Orientation to the sky is obtained from ground-based astrometry 
(USNO-A2.0 catalog, Monet 1998\nocite{Mon98}) with uncertainties in the field orientation $\pm 0\fdg05$.

\subsection{Assessing Reference Frame Residuals}
\begin{figure}
\epsscale{1.0}
\epsscale{0.6}
\plotone{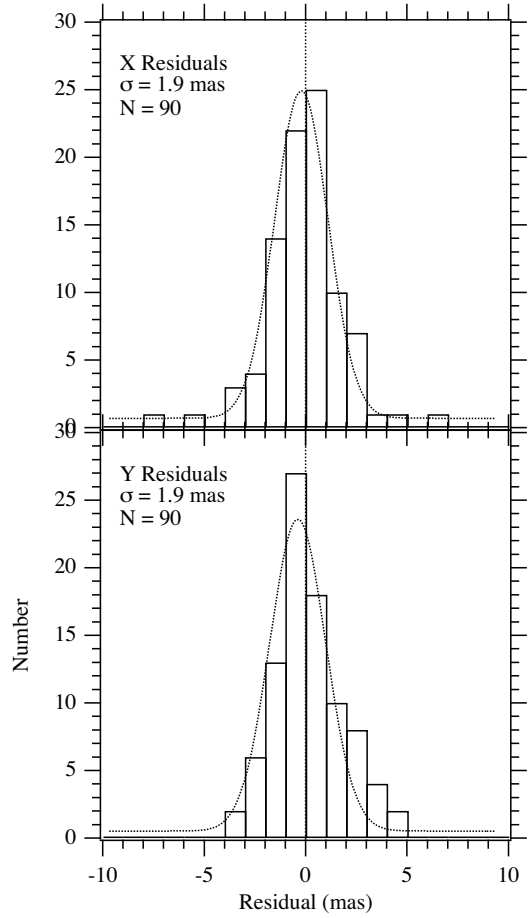}
\caption{Histograms of x and y residuals obtained from modeling the astrometric 
reference stars with equations 4 and 5. Distributions are fit
with gaussians whose $\sigma$'s are noted in the plots.} \label{fig-4}
\end{figure}

The Optical Field Angle Distortion calibration (\cite{McA97}) reduces as-built {\it HST} telescope and FGS 3 distortions with amplitude $\sim1\arcsec$ to below 2 mas over much of the FGS 3 field of regard. From histograms of the reference star astrometric residuals (Figure~\ref{fig-4}) we conclude that we have obtained satisfactory correction in the
region available at all {\it HST} rolls (an inscribed circle centered on the pickle-shaped FGS field of regard). The resulting reference frame 'catalog' in $\xi$ and $\eta$ standard coordinates (Table \ref{tbl-POS}) was determined
with	$<\sigma_\xi>= 1.0$	 and	$<\sigma_\eta> = 1.0$ mas.

To determine if there might be unmodeled - but possibly correctable -  systematic effects at the 1 mas level, we plotted the NGC 6853 reference frame X and Y residuals against a number of spacecraft, instrumental, and astronomical parameters. These included X, Y position within the pickle; radial distance from the pickle center; reference star V magnitude and B-V color; and epoch of observation.  We saw no obvious trends, other than an expected increase in positional uncertainty with reference star magnitude. 

\subsection{The Absolute Parallax of the NGC 6853 Central Star} \label{AbsPi}
In a quasi-Bayesian approach the reference star spectrophotometric absolute parallaxes were input as observations with associated errors, not as hardwired quantities known to infinite precision. 
We obtain for the NGC 6853 CS an absolute parallax $\pi_{abs} = 2.10 \pm0.48$ mas. We note that the formal error on this particular {\it HST} parallax is larger than we typically achieve with FGS 3. For eight objects in common with {\it HIPPARCOS} we obtain an average parallax precision, $\langle \sigma_{\pi}\rangle$ = 0.26 mas, with no statistically significant scale difference compared to {\it HIPPARCOS} (Benedict \etal 2002b and Benedict \etal 2002c). We attribute the larger than expected parallax error for the NGC 6853 CS to the less than optimum pattern of maximum positive and negative parallax factors seen in Table~\ref{tbl-LOO}.
However, our result agrees within the errors with the previous ground-based parallax measurement of the NGC 6853 CS (Harris \etal 1997), $\pi_{abs} = 2.63 \pm0.43$ mas. 
Parallaxes from {\it HST} and {\it USNO} and relative proper motion results from {\it HST} are collected in Table~\ref{tbl-SUM}. For the remainder of this paper we adopt as the absolute parallax of the NGC 6853 CS, $\pi_{abs} = 2.40 \pm0.32$ mas, the weighted average of these two completely independent parallax determinations.

\section{Discussion and Summary}
\subsection{FGS Photometry of the CS}
FGS 3 is a precision photometer, yielding relative photometry with 0.002 mag errors (Benedict \etal 1998\nocite{Ben98}). During each of the eight total observation sets we observed the NGC 6853 CS 6--7 times over approximately 33 minutes. No non-random variations
were noted within any one data set. The average $\sigma_V$ for the eight data sets was $\langle\sigma_V\rangle \sim$ 0.002 mag. However, statistically significant variations over the 2.6 year campaign duration are evident in Table~\ref{tbl-LOO}. Our coverage is too sparse to extract any periodic component to this variation.

\subsection{The Lutz-Kelker-Hanson Bias}

When using a trigonometric parallax to estimate the absolute
magnitude of a star, a correction should be made for the
Lutz-Kelker  bias (\cite{Lut73}) as modified by Hanson (1979).
Because of the galactic latitude and distance of NGC 6853, 
and the scale height of the
stellar population of which it is a member,
we use a uniform space density for determining
this LKH bias. The LKH bias is proportional to $(\sigma_{\pi}/\pi)^2$. Presuming that the CS belongs to the same class of object as RR Lyr (evolved Main Sequence stars), we scale the LKH correction determined for RR Lyr in Benedict \etal (2002a) and obtain LKH = -0.15 $\pm$ 0.07.
%Confirming this estimate, we obtain an LK bias correction, LK = $-0.12^{+0.18}_{-0.60}$ mag from \cite{Koe92}, table 1, with $\sigma_{\pi}/\pi = 0.13$ and our number of degrees of freedom in the solution, N= 95. The upper and lower bounds of the correction are 90\% confidence limits ($1.65\sigma$). Corresponding $1\sigma$ limits would yield LK = -0.12^{+0.09}_{-0.24}. 

\subsection{The Absolute Magnitude of the Central Star of NGC 6853}
Adopting for the NGC 6853 CS  
V= 13.98$\pm$ 0.03  and  the weighted average absolute parallax, $\pi_{abs} = 2.40 \pm0.32$ mas from Section \ref{AbsPi}, we determine a distance modulus, m-M = 8.10$\pm 0.15$. In Table \ref{tbl-AV} (Section \ref{AV}) we list a derived  per-star, per-unit 100 pc distance absorption, $\langle A$$_V$$\rangle$/100pc. Given the considerable scatter in that value, we adopt the average of the three stars nearest the central target (see Figure~\ref{fig-1}), ref-4, -5, and -8, $\langle$ A$_V\rangle$/100pc = 0.07$\pm$0.03. With this per-unit 100 pc $\langle$ A$_V\rangle$ and the measured distance to the NGC 6853 central star, d = 417$^{+49}_{-65}$ pc, we obtain a total absorption for the CS, A$_V^* = 0.30\pm0.06$. As a check we employ the formulation (Laney \& Stobie 1993), 
\beq
R  =  3.07 + 0.28\times (B-V)_0 + 0.04\times E(B-V)
\eeq
to obtain for the NGC 6853 CS, R = 2.99. From R = A$_V$/E(B-V) we then obtain E(B-V) = 0.10 and a reddening-corrected B-V = -0.34, a value one would expect for an object with T $\sim 10^5$ K. For further confirmation of this reddening value Ciardullo \etal (1999) \nocite{Cia99} obtain for NGC 6853 A$_V^*$ = 0.2$\pm$0.1 from an E(B-V), c relation, assuming a CS temperature of 10$^5$ K. With A$_V^* = 0.30\pm0.06$ we obtain M$_V = 5.43^{+0.28}_{-0.32}$, where we have included the LKH correction and its uncertainty and the 0.06 magnitude uncertainty in A$_V^*$ in quadrature. 

\subsection{A Central Star Radius}
To estimate a radius, $R_{*}$, for this star we require a distance, an absolute magnitude, an effective temperature, T$_{eff}$, and a bolometric correction (B.C.) . These quantities then yield a radius via differential comparison with the 
sun. Our parallax provides a distance, d = 417$^{+49}_{-65}$ pc and an absolute magnitude, M$_V = 5.43^{+0.28}_{-0.32}$. \cite{Nap99} has estimated T$_{eff} = 108,600 K \pm$ 6800 K from model atmosphere fits to the Balmer H$\delta$ and H$\epsilon$ absorption lines. 

For the B.C. we have two sources. Bergeron et al. (1995) tabulate B.C. up to T$_{eff} = 100,000$ K from a pure Hydrogen, log$~g = 8$ DA white dwarf model convolved with a V bandpass. A small extrapolation yields B.C. = -7.13. Flower (1996) provides bolometric corrections for normal stars up to T$_{eff} \sim 54,000K$. From Flower (1996), figure 4, the relationship between log T$_{eff}$ and B.C. is linear for  T$_{eff} >$ 25,000 K. Hotter stars lie on the Rayleigh-Jeans tail of the blackbody curve, where flux is roughly proportional to T$_{eff}$, not T$_{eff}^4$. A linear extrapolation yields B.C. = -7.03 for the NGC 6853 CS. We adopt B.C. = - 7.1 $\pm$ 0.2, where the error is dominated by the uncertainty in T$_{eff}$ and the poorly characterized behavior of the B.C. at these high temperatures.

We obtain a CS 
bolometric luminosity M$_{bol}$  = M$_V$ + B.C. = -1.67 $\pm 0.37$. $R_{*}$ follows from the expression
\beq
M_{bol}^{\sun} - M_{bol}^{*} = 10~log(T_{eff}^{*}/~T_{eff}^{\sun})+ 5~log(R_{*}/R_{\sun})
\eeq
where we assume for the Sun $M_{bol}^{\sun}=+4.75$ and $T_{eff}^{\sun}=5800 K$. 
We find $R_{*} = 0.05 \pm 0.02 R_{\sun}$. The sources of error for this radius are in the absolute magnitude (i.e., the parallax), the
bolometric correction, and the $T_{eff}^{*}$. 

A second way to obtain $R_{*}$ involves the V-band average flux, $H_V$, discussed in 
Bergeron et al. (1995). They list $H_V^{*}$ as a function of temperature for, again, a pure Hydrogen, $log~g = 8$ DA model. We obtain $H_V^{*}$ for $T_{eff}=108,600$ K by a small extrapolation from the highest temperature considered by Bergeron et al. (1995), $T_{eff}=100,000$ K. If we can determine an $H_V^{\sun}$, we can derive $R_{*}$ from
\beq
R_{*}^2 = (H_V^{\sun}/H_V^{*})10^{-0.4(M_V^{*}-M_V^{\sun})}
\eeq
In Benedict \etal (2000), where we estimated the radius of Feige 24, the exponent on $R_{*}$ was inadvertantly omitted, but not in the calculation. In that paper we obtained $H_V^{\sun}$ by convolving the Bessell (1990) V band response with the solar spectral distribution listed in Allen (1973). We calculated $H_V^{\sun} = 6.771\times10^5$ ergs cm$^{-2}$ s$^{-1}$ \AA$^{-1}$ str$^{-1}$.
We obtain for this CS with $T_{eff}=108,600 K$, $M_V^{*} = 5.43$ from our parallax, and $M_V^{\sun} = 4.82$ an $R_{*}=0.06 \pm 0.02 R_{\sun}$. As a final check we calculate $R_{*}$ from equation 8, but differentially with respect to our Feige 24 $M_V$ (Benedict \etal 2000, table 5), rather than $M_V^{\sun}$. This also yields $R_{*}=0.06 \pm 0.02 R_{\sun}$. Given that the approach relying directly on the B.C. and the approach utilizing $H_V$ yield $R_{*}$ values that agree within their errors, we adopt $R_{*}=0.055 \pm 0.02 R_{\sun}$. The error on this radius (2.75$\sigma$) cannot be further reduced by a weighted average of the results from the two approaches, because their errors are highly correlated, both having significant contributions from the uncertainties in $T_{eff}$ and M$_V^{*}$.

Comparing with the results presented in \cite{Pro98}, figure 3, and our radius for the hot white dwarf Feige 24, we find the CS of NGC 6853 to have a radius larger than any other white dwarf so far measured. 
On a log $g$ - log T$_{eff}$ diagram (Napiwotzki~1999\nocite{Nap99}, figure 2) the evolutionary track of a post-AGB star of a given mass traces a path at first of both increasing log $g$ and log T$_{eff}$. Once the star reaches the WD cooling track, log T$_{eff}$ then decreases, with a smaller rate of change in log $g$.  Having determined both T$_{eff}$ and log $g$ from line profile fitting, \cite{Nap99} estimates the mass of the NGC 6853 CS from such a diagram, obtaining $M = 0.56 \pm 0.01M_{\sun}$. A mass and radius uniquely determine a gravity, $g$,

\beq
g = MG /R^2
\eeq

where G is the gravitational constant. Our radius, $R_{*}=0.055 \pm 0.02 R_{\sun}$ and the \cite{Nap99} mass yield log $g$ = 6.7$\pm$0.4, in good agreement with the \cite{Nap99} line profile fitting value, log $g$ = 6.7$\pm$0.2. Our radius estimate is consistent with a stellar core not yet descended to the white dwarf cooling sequence.

\subsection{Summary}
{\it HST} astrometry yields an absolute trigonometric parallax for the NGC 6853 CS, $\pi_{abs} = 2.10 \pm0.48$ mas. A weighted average with a previous ground-based determination (Harris \etal 1997\nocite{Har97}) provides $\pi_{abs} = 2.40 \pm0.32$ mas. The higher precision resulting from the average of two independent parallax determinations requires a smaller LKH bias correction, -0.15$\pm$0.07 magnitude. Spectrophotometry of the astrometric reference stars local to NGC 6853 suggest an extinction for the CS, A$_V^*$ = 0.30$\pm$0.06. The dominant error terms in the resulting absolute magnitude, M$_V = 5.43^{+0.28}_{-0.32}$, are the parallax and the uncertainty in the amount of extinction for the CS itself. Two methods for estimating the radius of the CS yield $R_{*}=0.055 \pm 0.02R_{\sun}$, suggesting a star still above the white dwarf cooling track.

\acknowledgments

We thank Conard Dahn for his careful and critical refereeing of this paper. 
Support for this work was provided by NASA through grant NAG5-1603 from the Space 
Telescope 
Science Institute, which is operated
by AURA, Inc., under
NASA contract NAS5-26555. These results are based partially on observations 
obtained with the
Apache Point Observatory 3.5 m telescope, which is owned and operated by
the Astrophysical Research Consortium. This publication makes use of data products 
from the Two Micron All Sky Survey (2MASS),
which is a joint project of the University of Massachusetts and the Infrared 
Processing
and Analysis Center/California Institute of Technology, funded by NASA and the 
NSF. 
This research has made use of the SIMBAD database, operated at CDS, Strasbourg, 
France; the NASA/IPAC Extragalactic Database (NED) which is operated by
            JPL, California Institute of Technology, under contract with the NASA;  
and NASA's Astrophysics
Data System Abstract Service. 

\clearpage

% Now comes the reference list.  

%

\clearpage

\begin{deluxetable}{llll}
\tablewidth{4in}
\tablecaption{NGC6853 Log of Observations\label{tbl-LOO}}
\tablehead{\colhead{Set}&
\colhead{mJD}&\colhead{Roll (\arcdeg) \tablenotemark{a}}&
\colhead{V\tablenotemark{b}} }
\startdata
1&49985.60531&101.91&13.985$\pm$0.004\nl
2&50021.33038&103.9&13.979$\pm$0.002\nl
3&50172.92498&268.6&14.015$\pm$0.003\nl
4&50178.95797&270.4&14.016$\pm$0.002\nl
5&50208.6424&286.6&14.012$\pm$0.002\nl
6&50233.84447&286.6&14.026$\pm$0.002\nl
7&50569.61249&286.0&13.994$\pm$0.003\nl
8&50932.79954&286.0&14.007$\pm$0.003\nl
\enddata
\tablenotetext{a}{Spacecraft roll as defined in Chapter 2, FGS Instrument Handbook 
(Nelan, 2001) }
\tablenotetext{b}{Average of 6 to 7 observations at each epoch. Errors are 
internal, not external.
}
\end{deluxetable}

\begin{deluxetable}{llllllll}
\tablewidth{0in}
\tablecaption{CS  and Reference Star Data    \label{tbl-POS}}
\tablehead{\colhead{ID}&
\colhead{$\xi$ \tablenotemark{a}} &
\colhead{$\eta$ \tablenotemark{a}} &
\colhead{$\mu_x$ \tablenotemark{b}} &
\colhead{$\mu_y$ \tablenotemark{b}} }
\startdata\nl
CS\tablenotemark{c}&0.0000$\pm$0.0006&0.0000$\pm$0.0005&0.0069$\pm$0.0005&0.0179$\pm$0.0004\nl
ref-2&-62.9877$\pm$0.0007&-90.3490$\pm$0.0008&-
0.0013$\pm$0.0010&0.0052$\pm$0.0007\nl
ref-3&-154.2862$\pm$0.0008&-20.5996$\pm$0.0009&0.0028$\pm$0.0026&  -
0.0112$\pm$0.0020\nl
ref-4&-13.2770$\pm$0.0007 &48.4715$\pm$0.0007 & -0.0066$\pm$0.0026 &0.0098$\pm$ 
&0.0021\nl
ref-5&15.5963$\pm$0.0008&34.4350$\pm$0.0007&0.0058$\pm$0.0031&0.0123$\pm$0.0025\nl
ref-6&93.1016$\pm$0.0008&89.0745$\pm$0.0008&0.0004$\pm$0.0006&0.0020$\pm$0.0006\nl
ref-7&164.5535$\pm$0.0010&51.1593$\pm$0.0011&0.0026$\pm$0.0024&-
0.0107$\pm$0.0019\nl
ref-8&67.3571$\pm$0.0008&40.5966$\pm$0.0008&-
0.0064$\pm$0.0023&0.0048$\pm$0.0019\nl
\enddata
\tablenotetext{a}{$\xi$ and $\eta$ are relative positions in arcseconds}
\tablenotetext{b}{$\mu_x$ and $\mu_y$ are relative motions in arcsec yr$^{-1}$ }
\tablenotetext{c}{RA = 19 59 36.38 Dec = 22 43 16.0, J2000, epoch = mJD 51486.092. CS position from 2MASS.}
\end{deluxetable}

\begin{deluxetable}{llllllll}
\tablewidth{0in}
\tablecaption{Visible Photometry
\label{tbl-VIS}}
\tablehead{\colhead{ID}&
\colhead{V} &\colhead{B-V} &
\colhead{V-R} &
\colhead{V-I} &
\colhead{V-K} }
\startdata
CS& 13.98$\pm$0.03&-0.24$\pm$0.04&-0.46$\pm$0.04&-0.08$\pm$0.04& \nl
ref-2&15.41$\pm$0.06&2.2$\pm$0.2&1.11$\pm$0.08&2.19$\pm$0.08&3.91$\pm$0.06\nl
ref-3&11.66$\pm$0.02&1.87$\pm$0.03&1.47$\pm$0.03&2.57$\pm$0.03&4.52$\pm$0.03\nl
ref-4&14.94$\pm$0.03&0.76$\pm$0.07&0.59$\pm$0.04&1.21$\pm$0.04&1.98$\pm$0.04\nl
ref-5&15.45$\pm$0.04&0.92$\pm$0.07&0.55$\pm$0.05&1.18$\pm$0.05&2.03$\pm$0.05\nl
ref-6&14.10$\pm$0.03&1.95$\pm$0.07&1.58$\pm$0.04&2.66$\pm$0.04&4.63$\pm$0.03\nl
ref-7&13.71$\pm$0.03&0.66$\pm$0.06&0.49$\pm$0.04&1.05$\pm$0.04&&\nl
ref-8&14.69$\pm$0.03&1.98$\pm$0.07&1.12$\pm$0.04&2.13$\pm$0.04&4.07$\pm$0.04\nl
\enddata
\end{deluxetable}

\begin{deluxetable}{llllllll}
\tablewidth{0in}
\tablecaption{Astrometric Reference Stars Near-IR and Washington-DDO
Photometry
  \label{tbl-IR}}
\tablehead{\colhead{ID}&
\colhead{K} &
\colhead{J-H} &
\colhead{H-K} &
\colhead{M-T$_2$} &
\colhead{M-51} }
\startdata
ref-2&11.50$\pm$0.02&0.74$\pm$0.04&0.15$\pm$0.04&&&&\nl
ref-3&7.14$\pm$0.02&0.89$\pm$0.03&0.23$\pm$0.03&2.51$\pm$0.01&-0.17$\pm$0.01\nl
ref-4&12.96$\pm$0.03&0.35$\pm$0.04&0.13$\pm$0.04&1.34$\pm$0.10&-0.01$\pm$0.14\nl
ref-5&13.42$\pm$0.03&0.42$\pm$0.04&0.10$\pm$0.05&0.56$\pm$0.09&-0.02$\pm$0.1\nl
ref-6&9.47$\pm$0.02&0.81$\pm$0.03&0.25$\pm$0.03&&&&\nl
ref-7&&&&0.99$\pm$0.03&0.10$\pm$0.03\nl
ref-8&10.62$\pm$0.03&0.80$\pm$0.04&0.24$\pm$0.04&1.97$\pm$0.08&-0.09$\pm$0.11\nl
\enddata
\end{deluxetable}

\begin{center}
\begin{deluxetable}{llllllll}
\tablewidth{0in}
\tablecaption{Reference Star A$_V$ from Spectrophotometry  \label{tbl-AV}}
\tablehead{  \colhead{ID}&
\colhead{A$_V$(V-I)}&   \colhead{A$_V$(V-R)}&  \colhead{A$_V$(V-K)} &  
\colhead{A$_V$(J-K)}&
\colhead{$\langle $A$_V\rangle$}&
\colhead{$\langle $A$_V\rangle$/100pc}}
\startdata
ref-2&1.64&2.02&1.68&1.68&1.75$\pm$0.10&0.05$\pm$0.01\nl
ref-3&2.44&2.09&1.59&2.09&2.05$\pm$0.20&0.29$\pm$0.03\nl
ref-4&0.29&1.13&0.51&0.64&0.64$\pm$0.21&0.13$\pm$0.04\tablenotemark{*}\nl
ref-5&0.04&0.07&0.51&0.81&0.36$\pm$0.21&0.04$\pm$0.02\tablenotemark{*}\nl
ref-6&3.89&3.07&2.43&2.73&3.03$\pm$0.36&0.26$\pm$0.03\nl
ref-7&0.55&1.27&0.00&0.00&0.91$\pm$0.36&0.10$\pm$0.04\nl
ref-8&1.37&1.61&1.44&1.97&1.60$\pm$0.16&0.05$\pm$0.01\tablenotemark{*}\nl
\enddata
\tablenotetext{*}{Stars used to estimate $\langle $A$_V\rangle$/100pc for CS}
\end{deluxetable}
\end{center}

\begin{deluxetable}{llllll}
\tablewidth{0in}
\tablecaption{Astrometric Reference Star Spectral Classifications and
Spectrophotometric Parallaxes \label{tbl-SPP}}
\tablehead{\colhead{ID}& \colhead{Sp. T.}&
\colhead{V} & \colhead{M$_V$} & \colhead{A$_V$} &
\colhead{$\pi_{abs}$}(mas) } 
\startdata
ref-2&K0 III&15.41&0.7$\pm$0.4&1.8&0.3$\pm$0.1\nl
ref-3&K3 III&11.66&0.3$\pm$0.4&2.1&1.4$\pm$0.2\nl
ref-4&G3 V&14.94&5.9$\pm$0.4&0.6&2.1$\pm$0.4\nl
ref-5&G5 V&15.45&5.1$\pm$0.4&0.4&1.2$\pm$0.5\nl
ref-6&K0 III&14.10&0.7$\pm$0.4&3.0&0.8$\pm$0.2\nl
ref-7&F4 V&13.71&3.3$\pm$0.4&0.9&1.3$\pm$0.2\nl
ref-8&K2 III&14.69&0.5$\pm$0.4&1.6&0.3$\pm$0.1\nl
\enddata
\end{deluxetable}

\begin{center}
\begin{deluxetable}{ll}
\tablecaption{NGC 6853  Parallax and Proper Motion \label{tbl-SUM}}
\tablewidth{0in}
\tablehead{\colhead{Parameter} &  \colhead{ Value }}
\startdata
{\it HST} study duration  &2.59 y\nl
number of observation sets    &   8 \nl
ref. stars $ <V> $ &  $14.28 $  \nl
ref. stars $ <B-V> $ &  $1.47 $ \nl
\nl
{\it HST} Absolute Parallax   & 2.10  $\pm$  0.48   mas\nl
USNO Absolute Parallax   & 2.63  $\pm$  0.43   mas\nl
Weighted Average Absolute Parallax& 2.40  $\pm$  0.32   mas\nl
{\it HST} Proper Motion  &19.2  $\pm$  1.1 mas y$^{-1}$ \nl
 \indent in pos. angle & 21\arcdeg  $\pm$ 1\arcdeg \nl
\enddata
\end{deluxetable}
\end{center}

\begin{center}
\begin{deluxetable}{lll}
\tablecaption{NGC 6853 Central Star Astrophysical Quantities \label{tbl-6}}
\tablewidth{0in}
\tablehead{\colhead{Parameter} &  \colhead{Value} &  \colhead{Source}}
\startdata
$ V $    &   13.98 $\pm$ 0.03 & this paper, Table~\ref{tbl-VIS}\\
\bv & -0.24 $\pm$ 0.04 & this paper, Table~\ref{tbl-VIS} \\
d & 417$^{+49}_{-65}$ pc & this paper\\
$A_V^*$ &$ 0.30 \pm 0.06 $ & d \& $\langle $A$_V\rangle$/100pc for ref-4, 5, 8 \\
m-M  & $8.10 \pm 0.15$ & this paper\\
LKH Bias & -0.15 $\pm$ 0.07 & this paper\\
$M_V$ & $5.43^{+0.28}_{-0.32}$ & m-M, $A_V$, LKH Bias \\
$T_{eff}^{*}$ & $108,600 \pm 6,800 K$ &  \cite{Nap99}\\
B.C. & $-7.1 \pm 0.20$ & Flower (1996), \cite{Ber95} \\
$M_{bol}^{*}$&  $-1.67 \pm 0.37$& = $M_V$ +B.C. \\
$R_{*} $ & $ 0.055 \pm 0.02 R_{\sun}$& this paper\\
${\cal M}_{*}$ & $ 0.56\pm 0.01 {\cal M}_{\sun}$ &  \cite{Nap99}\\
log $g$ & 6.7 $\pm$ 0.2, 6.7 $\pm$ 0.4& \cite{Nap99}, this paper \\
\enddata
\end{deluxetable}
\end{center}

% And finally, we must deal with the figures.  There are three figures
% associated with this manuscript; two figures are Encapsulated
% PostScript (EPS) files.  The third figure is a grey scale figure that does
% not exist in EPS form.
%
% Authors have three options for including figure information within a
% manuscript.  Not all the options may be acceptable by the target Journal 
%- be
% sure to look at the appropriate submission instructions, electronic or
% otherwise.
%

\clearpage

\clearpage

\clearpage
\clearpage

\clearpage

\end{document}